\documentclass[%
 reprint,
 superscriptaddress,
 amsmath,amssymb,
 aps,
 prd,
 nolongbibliography,
]{revtex4-2}

\usepackage{xcolor}
\usepackage{graphicx}
\usepackage{dcolumn}
\usepackage{multirow}
\usepackage{float}
\usepackage{bm}
\usepackage[mathlines]{lineno}

\newcommand{\epem}{e^{+}e^{-}}
\newcommand{\ket}[1]{\left|#1\right\rangle}

\newcommand{\dz}{D^{0}}

\newcommand{\DDbar}{D\bar{D}}

\newcommand{\km}{K^{-}}
\newcommand{\pip}{\pi^{+}}
\newcommand{\pim}{\pi^{-}}
\newcommand{\piz}{\pi^{0}}

\usepackage[colorlinks,linkcolor=blue,anchorcolor=blue,citecolor=blue,urlcolor=blue]{hyperref}

\hypersetup{hypertex=true,
	colorlinks=true,
	linkcolor=blue,
	anchorcolor=blue,
	citecolor=blue,
	urlcolor=blue}

\makeatletter
\g@addto@macro\frontmatter@setup{\parskip=0pt}
\makeatother

\begin{document}

\preprint{APS/123-QED}

\title{Prospects for exploring quantum-correlated neutral $D\bar{D}$ decays at STCF}

\author{Ying-Hao~Wang}
\affiliation{Lanzhou University, Lanzhou 730000, People's Republic of China}
\author{Yi-Feng Wu}
\email[Corresponding author: ]{wyifeng2024@lzu.edu.cn}
\affiliation{Lanzhou University, Lanzhou 730000, People's Republic of China}
\author{Pei-Rong~Li}
\affiliation{Lanzhou University, Lanzhou 730000, People's Republic of China}
\affiliation{MOE Frontiers Science Center for Rare Isotopes, Lanzhou University, Lanzhou 730000, People's Republic of China}
\author{Yu~Zhang}%
\email[Corresponding author: ]{yuzhang@usc.edu.cn}
\affiliation{University of South China, Hengyang 421001, People's Republic of China}
\author{Xiao-Rui~Lyu}%
\email[Corresponding author: ]{xiaorui@ucas.ac.cn}
\affiliation{University of Chinese Academy of Science, Beijing 100049, People's Republic of China}

\date{\today}

\begin{abstract}
We investigate the prospects for determining charm mixing, indirect $C\!P$ violation and hadronic parameters in charm decays with quantum-correlated neutral charm mesons at the proposed Super Tau-Charm Facility. The study is based on time-integrated double tag rates of coherent $\DDbar$ pairs produced at $\sqrt{s}=3.773$ and $4.030$ GeV, 
from $\epem\to D^{(*)}\bar{D}^{(*)}$ processes.
Here, one $D$ is tagged with the quasi flavour-specific decays $D\to K^-\pi^+\pi^+\pi^-$, $K^-\pi^+\pi^0$, and $K^-\pi^+$, while the other $\bar{D}$ is detected with the final states in flavour-specific mode, $C\!P$ eigenstate mode, or self-conjugate mode. The expected sensitivities are evaluated assuming an integrated luminosity of $1~{\rm ab}^{-1}$ at each energy point, demonstrating the capability on precise determination of the charm mixing parameters, the indirect $C\!P$ violation parameters, the strong-phase differences, and the coherence factors. The constraints of the hadronic-parameters provide unique inputs to the CKM angle $\gamma$ fit, while the measurements of charm mixing and indirect $C\!P$ violation parameters are complementary to the time-dependent measurements in the Belle II and LHCb experiments.
\end{abstract}

\makeatletter
\let\compose@footnotes@two\compose@footnotes@one
\makeatother
\maketitle


\section{Introduction}
\label{sec:introduction}

Neutral-meson mixing has been established in the $K^{0}\bar{K}^{0}$, $B^{0}\bar{B}^{0}$, $B_s^{0}\bar{B}_s^{0}$ and $D^{0}\bar{D}^{0}$ systems~\cite{PDG:2026}.
The $D^{0}\bar{D}^{0}$ system is of particular interest, as it provides the only observed example of neutral-meson mixing in the up-type quark sector.  
An initially produced flavour eigenstate, $D^0$ or $\bar{D}^0$, can evolve into its antiparticle before decay. In the Standard Model this transition is governed by short-distance weak amplitudes and long-distance hadronic intermediate states, and reliable predictions are limited by non-perturbative QCD effects~\cite{mixing_and_CPV}.  Many extensions of the Standard Model can modify the mixing amplitude or introduce additional weak phases. Precision measurements of charm mixing and indirect $C\!P$ violation therefore serve as probes of new physics~\cite{Isidori:2010kg}.

The physical mass eigenstates can be expressed as superpositions of the flavour eigenstates
\begin{eqnarray}
    \ket{D_1} &= p\ket{D^0} + q\ket{\bar{D}^0}, \nonumber\\
    \ket{D_2} &= p\ket{D^0} - q\ket{\bar{D}^0}.\label{eq1}
\end{eqnarray} 
The complex coefficients $p$ and $q$ satisfy the normalization condition $|p|^2+|q|^2=1$. In the convention $C\!P\ket{D^0}=\ket{\bar{D}^0}$ and $C\!P\ket{\bar{D}^0}=\ket{D^0}$, the states $\ket{D_1}$ and $\ket{D_2}$ are $C\!P$-even and $C\!P$-odd, respectively, in the absence of $C\!P$ violation. 
Neglecting the $C\!P$ violation in decay, the indirect $C\!P$ violation is conventionally parameterized by a deviation of $|q/p|$ from unity and by the weak phase $\phi\equiv\arg(q/p)$.
Usually, the mixing parameters are defined by:
\begin{equation}
   \begin{aligned}
      x\equiv \frac{m_{1}-m_{2}}{\Gamma},~y\equiv \frac{\Gamma_{1}-\Gamma_{2}}{2\Gamma},
    \end{aligned}
\end{equation}
where the mass and width of the $D_{1(2)}$ state are denoted by $m_{1(2)}$ and $\Gamma_{1(2)}$, respectively. The average width is defined as $\Gamma = (\Gamma_{1}+\Gamma_{2})/2$.
Evidence for charm mixing was first reported by BABAR and Belle in 2007~\cite{BaBar:ddbarmix_eviden,BELLE:ddbarmix_eviden}, and was subsequently supported by CDF~\cite{CDF:ddbarmix_eviden}.  
LHCb later observed oscillations using the wrong-sign decay $\dz\to K^+\pi^-$~\cite{LHCb:first_mix}. Measurements in the same mode by CDF and Belle provided independent confirmation~\cite{CDF:mixing,Belle:mixing}.  
More recently, an updated combination of LHCb measurements has provided a simultaneous determination of the charm mixing and indirect $C\!P$ violation parameters in the neutral-charm system~\cite{LHCb:2024yxi}. The current HFLAV global fit allowing for $C\!P$ violation gives $x=(0.405\pm0.043)\%$, $y=(0.636\pm0.024)\%$, $|q/p|=0.983^{+0.015}_{-0.014}$, and $\phi=(-1.51^{+1.03}_{-1.06})^\circ$~\cite{HFLAV:CKM2025}.

At charm threshold, the time-integrated double tag (DT) decay rates of quantum-correlated $\DDbar$ pairs provide complementary constraints on charm mixing, indirect $C\!P$ violation, and the hadronic parameters of $D$ decays.
In $\epem$ annihilation through a single virtual photon with $J^{PC}=1^{--}$, conservation of $C$-parity determines the coherent state of the $\DDbar$ system. For a $\DDbar$ pair produced in association with $m$ neutral pions and $n$ photons, the charge-conjugation eigenvalue of the $\DDbar$ system is $C=(-1)^{n+1}$.
Consequently, both $C$-odd and $C$-even quantum-correlated $\DDbar$ samples can be selected through processes involving different associated neutral particles. Their time-integrated DT rates depend on combinations of the charm mixing parameters $x$ and $y$, the indirect $C\!P$ violation parameters $|q/p|$ and $\phi$, and hadronic parameters such as the strong-phase differences and coherence factors. Threshold measurements can therefore be used both to determine the hadronic parameters and to constrain charm mixing and indirect $C\!P$ violation without relying on decay-time information.
BESIII has measured the strong-phase difference in $D\to K^-\pi^+$~\cite{BESIII:KpiStrongPhase} and the hadronic parameters of $D\to K^-\pi^+\pi^+\pi^-$ and $D\to K^-\pi^+\pi^0$~\cite{8fb-1} using quantum-correlated $\DDbar$ samples. These hadronic parameters are important external inputs to charm mixing measurements and determinations of the CKM angle $\gamma$. With the larger data samples expected at LHCb and Belle II, the precision of charm mixing measurements and determinations of $\gamma$ is expected to improve~\cite{LHCb:UpgradeII,BelleII:PhysicsBook}. The uncertainties associated with the external hadronic inputs may consequently become more significant. Larger data samples at charm threshold would improve the determination of the hadronic parameters and provide constraints on charm mixing and indirect $C\!P$ violation that are complementary to those from time-dependent measurements.

The proposed Super Tau-Charm Facility (STCF) is a symmetric electron-positron collider designed to operate over the center-of-mass energy range of $2$--$7$ GeV, with a peak luminosity of at least $0.5\times10^{35}\,{\rm cm}^{-2}{\rm s}^{-1}$~\cite{STCF:CDR}. Its high luminosity in the open-charm threshold region is expected to provide large samples of quantum-correlated neutral-$D$-meson pairs~\cite{Cheng:2022tog}. 
In this work, we consider coherent $\DDbar$ samples produced at $\sqrt{s}=3.773$ and $4.030$ GeV, which provide complementary $C$-odd and $C$-even configurations. We investigate the prospects for determining the charm mixing parameters, the indirect $C\!P$ violation parameters and the hadronic parameters of the quasi-flavour-specific decays $D\to K^-\pi^+\pi^+\pi^-$, $D\to K^-\pi^+\pi^0$, and $D\to K^-\pi^+$ using time-integrated DT decay rates. 
The expected sensitivities are evaluated assuming an integrated luminosity of $1~{\rm ab}^{-1}$ at each energy point.

The remainder of this paper is organized as follows. The analysis strategy, including the DT method, the signal and tag modes, and the formalism for the time-integrated rates of quantum-correlated decays, is presented in Sec.~\ref{sec:analysis-strategy}. The Monte Carlo simulation and event selection are discussed in Sec.~\ref{sec:simulation-selection}, followed by the simultaneous-fit procedure in Sec.~\ref{sec:fit-procedure}. Section~\ref{sec:expected-sensitivities} presents the expected sensitivities to the hadronic, charm mixing, and indirect $C\!P$ violation parameters, together with the impact of the hadronic-parameter inputs on future measurements of the CKM angle $\gamma$. A summary is given in Sec.~\ref{sec:summary}.

\section{Analysis strategy}
\label{sec:analysis-strategy}

\subsection{DT strategy}

To evaluate the expected sensitivities to the charm mixing, indirect $C\!P$ violation, and hadronic parameters, we employ the DT method, in which both $D$ mesons in a quantum-correlated $\DDbar$ pair are reconstructed. 
The decay mode of interest is denoted by $f$ and referred to as the signal mode, while the decay mode of the other $D$ meson is denoted by $g$ and referred to as the tag mode. 
At $\sqrt{s}=3.773$~GeV, the $\DDbar$ pair is produced directly with $C=-1$. 
For $D^{(*)}\bar{D}^{(*)}$ production at $\sqrt{s}=4.030$~GeV, the $C$ eigenvalue of the selected pair is determined by the accompanying neutral particles. Complementary $C$-odd and $C$-even DT samples are therefore provided by the two energies.

The signal side includes the three decay modes $D\to K^-\pi^+\pi^+\pi^-$, $D\to K^-\pi^+\pi^0$, and $D\to K^-\pi^+$.
The tag modes are divided into like-sign (LS) and opposite-sign (OS) flavour tags, $C\!P$-even and $C\!P$-odd tags, and a self-conjugate mode $D\to K_S^0\pi^+\pi^-$, as summarized in Table~\ref{tab:tag-modes}.
For the quasi-$C\!P$ eigenstate $\pi^+\pi^-\pi^0$, the eigenvalue is replaced by the dilution factor $2F_+-1$, where $F_+$ is the $C\!P$-even fraction. The value $F_+^{\pi^+\pi^-\pi^0}=0.9406\pm0.0055_{\rm stat}\pm0.0033_{\rm syst}$, measured by the BESIII Collaboration, is used in this study~\cite{BESIII:FplusPipipi0}.

\begin{table}[H]
\caption{Tag modes used with the three signal decays. Charge-conjugate modes are implied.}
\label{tab:tag-modes}
\centering
\small
\setlength{\tabcolsep}{2pt}
\renewcommand{\arraystretch}{1.08}
\begin{tabular}{lcc}
\hline
\multirow{2}{*}{Flavour} & LS
& \parbox[t]{0.62\columnwidth}{\centering $K^-\pi^+\pi^+\pi^-$, $K^-\pi^+\pi^0$, $K^-\pi^+$} \\
& OS 
& \parbox[t]{0.62\columnwidth}{\centering $K^+\pi^-\pi^-\pi^+$, $K^+\pi^-\pi^0$, $K^+\pi^-$} \\
\hline
\multirow{2}{*}{$C\!P$} & Even
& \parbox[t]{0.62\columnwidth}{\centering $\pi^+\pi^-\pi^0$, $K^+K^-$, $\pi^+\pi^-$, $K_S^0\pi^0\pi^0$} \\
& Odd
& \parbox[t]{0.62\columnwidth}{\centering $K_S^0\pi^0$, $K_S^0\eta$, $K_S^0\omega$, $K_S^0\eta'$} \\
\hline
\multicolumn{2}{l}{Self-conjugate}
& \parbox[t]{0.62\columnwidth}{\centering $K_S^0\pi^+\pi^-$} \\
\hline
\end{tabular}
\end{table}

\subsection{Time-integrated DT rates}
The normalized quantum-correlated state of a $\DDbar$ pair with charge-conjugation eigenvalue $C$ is
\begin{equation}
\ket{\Psi_C}
=
\frac{1}{\sqrt{2}}
\left[
\ket{D^0}_1\ket{\bar{D}^0}_2
+C\ket{\bar{D}^0}_1\ket{D^0}_2
\right].
\label{eq:coherent-state}
\end{equation}
Here, $C=-1$ and $C=+1$ denote the $C$-odd and $C$-even states, respectively, while the subscripts distinguish the two $D$ mesons.
The joint decay amplitude is constructed from this state.
The two components of the quantum-correlated state contribute coherently to the joint decay amplitude, producing interference terms in the DT rates. The rates therefore cannot be factorized into products of two branching fractions, and the interference terms provide sensitivity to charm mixing, indirect $C\!P$ violation, and the relative magnitudes and phases of the signal- and tag-side amplitudes.
For a unified treatment of binned and unbinned measurements, $f_i$ and $g_j$ denote the $i$th and $j$th phase-space bins of the signal mode and tag mode, respectively. In the unbinned case, the full phase space is treated as a single bin, and the bin indices are omitted. The decay amplitudes for $D^0 \to f$ and $\bar D^0 \to f$ at a phase-space point $\Phi$ are denoted by $A_f(\Phi)$ and $\bar A_f(\Phi)$, respectively. The relevant quantities are defined as
\begin{minipage}{\columnwidth}
\begin{equation*}
\begin{aligned}
P_f^i&=\int_i|A_f(\Phi)|^2\,\mathrm{d}\Phi,\\
\bar P_f^i&=\int_i|\bar A_f(\Phi)|^2\,\mathrm{d}\Phi,\\
Z_f^i&=\int_i A_f(\Phi)\bar A_f^*(\Phi)\,\mathrm{d}\Phi,
\end{aligned}
\end{equation*}
\end{minipage}
with analogous definitions for $P_g^j$, $\bar P_g^j$, and $Z_g^j$.

In the absence of direct $C\!P$ violation in the decay amplitudes, the above definitions imply the following relations between $C\!P$-conjugate bins:
\begin{equation*}
P_{\bar f}^{\,i}=\bar P_f^i,\qquad
\bar P_{\bar f}^{\,i}=P_f^i,\qquad
Z_{\bar f}^{\,i}=(Z_f^i)^*,
\end{equation*}
and similarly for $g$. The time-integrated DT rates for the four charge combinations $(f_i,g_j)$, $(f_i,\bar g_j)$, $(\bar f_i,g_j)$, and $(\bar f_i,\bar g_j)$ are obtained by integrating the squared joint decay amplitude over both decay times. Retaining terms through second order in $x$ and $y$, the rates are expressed as follows.
\begin{widetext}
\small
\begin{equation}
\begin{aligned}
\Gamma_C(f_i;g_j)
&\propto \big[2-(2+C)(x^2-y^2)\big]\Big[P_f^i\bar P_g^j+\bar P_f^iP_g^j+2C\operatorname{Re}\!\left(Z_f^i(Z_g^j)^*\right)\Big]
\\
&\mathrel{+}(2+C)(x^2+y^2)\Big[\left|\frac pq\right|^2P_f^iP_g^j+\left|\frac qp\right|^2\bar P_f^i\bar P_g^j+2C\operatorname{Re}\!\left(e^{-2\mathrm{i}\phi}Z_f^iZ_g^j\right)\Big]
\\
&\mathrel{-}2(1+C)\operatorname{Re}\Bigg\{(y+\mathrm{i}x)\Bigg[\left|\frac pq\right|e^{-\mathrm{i}\phi}\left(P_f^iZ_g^j+C P_g^jZ_f^i\right)+\left|\frac qp\right|e^{\mathrm{i}\phi}\left(\bar P_f^i(Z_g^j)^*+C\bar P_g^j(Z_f^i)^*\right)\Bigg]\Bigg\}
\\[1.4ex]
\Gamma_C(f_i;\bar g_j)
&\propto \big[2-(2+C)(x^2-y^2)\big]\Big[P_f^iP_g^j+\bar P_f^i\bar P_g^j+2C\operatorname{Re}\!\left(Z_f^iZ_g^j\right)\Big]
\\
&\mathrel{+}(2+C)(x^2+y^2)\Big[\left|\frac pq\right|^2P_f^i\bar P_g^j+\left|\frac qp\right|^2\bar P_f^iP_g^j+2C\operatorname{Re}\!\left(e^{-2\mathrm{i}\phi}Z_f^i(Z_g^j)^*\right)\Big]
\\
&\mathrel{-}2(1+C)\operatorname{Re}\Bigg\{(y+\mathrm{i}x)\Bigg[\left|\frac pq\right|e^{-\mathrm{i}\phi}\left(P_f^i(Z_g^j)^*+C\bar P_g^jZ_f^i\right)+\left|\frac qp\right|e^{\mathrm{i}\phi}\left(\bar P_f^iZ_g^j+C P_g^j(Z_f^i)^*\right)\Bigg]\Bigg\}
\\[1.4ex]
\Gamma_C(\bar f_i;g_j)
&\propto \big[2-(2+C)(x^2-y^2)\big]\Big[P_f^iP_g^j+\bar P_f^i\bar P_g^j+2C\operatorname{Re}\!\left(Z_f^iZ_g^j\right)\Big]
\\
&\mathrel{+}(2+C)(x^2+y^2)\Big[\left|\frac pq\right|^2\bar P_f^iP_g^j+\left|\frac qp\right|^2P_f^i\bar P_g^j+2C\operatorname{Re}\!\left(e^{-2\mathrm{i}\phi}(Z_f^i)^*Z_g^j\right)\Big]
\\
&\mathrel{-}2(1+C)\operatorname{Re}\Bigg\{(y+\mathrm{i}x)\Bigg[\left|\frac pq\right|e^{-\mathrm{i}\phi}\left(\bar P_f^iZ_g^j+C P_g^j(Z_f^i)^*\right)+\left|\frac qp\right|e^{\mathrm{i}\phi}\left(P_f^i(Z_g^j)^*+C\bar P_g^jZ_f^i\right)\Bigg]\Bigg\}
\\[1.4ex]
\Gamma_C(\bar f_i;\bar g_j)
&\propto \big[2-(2+C)(x^2-y^2)\big]\Big[P_f^i\bar P_g^j+\bar P_f^iP_g^j+2C\operatorname{Re}\!\left(Z_f^i(Z_g^j)^*\right)\Big]
\\
&\mathrel{+}(2+C)(x^2+y^2)\Big[\left|\frac pq\right|^2\bar P_f^i\bar P_g^j+\left|\frac qp\right|^2P_f^iP_g^j+2C\operatorname{Re}\!\left(e^{-2\mathrm{i}\phi}(Z_f^i)^*(Z_g^j)^*\right)\Big]
\\
&\mathrel{-}2(1+C)\operatorname{Re}\Bigg\{(y+\mathrm{i}x)\Bigg[\left|\frac pq\right|e^{-\mathrm{i}\phi}\left(\bar P_f^i(Z_g^j)^*+C\bar P_g^j(Z_f^i)^*\right)+\left|\frac qp\right|e^{\mathrm{i}\phi}\left(P_f^iZ_g^j+C P_g^jZ_f^i\right)\Bigg]\Bigg\}.
\end{aligned}
\label{eq:time-integrated-DT-rates}
\end{equation}
\end{widetext}


For the $D\to K^-\pi^+\pi^+\pi^-$ and $D\to K^-\pi^+\pi^0$ modes, the hadronic parameters in $i$-th bin are defined by
\begin{equation*}
(r_f^i)^2=\frac{\bar P_f^i}{P_f^i},\qquad
R_f^i e^{\mathrm{i}\delta_f^i}=
\frac{Z_f^i}{\sqrt{P_f^i\bar P_f^i}},
\end{equation*}
where $r_f^i$ is the amplitude ratio, $R_f^i$ is the coherence factor, and $\delta_f^i=\arg Z_f^i$ is the average strong-phase difference in bin $i$, defined by the phase of the bin-integrated interference term. The tag-side quantities are defined analogously for $g$. These definitions also apply to modes analyzed without binning, with integration over the full phase space and the bin indices omitted, yielding $r_f$, $R_f$, and $\delta_f$.
The phase-space binning scheme for $D\to K^-\pi^+\pi^+\pi^-$ is adopted from Ref.~\cite{Evans:K3piBinning}.
The self-conjugate tag mode $D\to K_S^0\pi^+\pi^-$ is analyzed using the equal-binning scheme. Each half of the dalitz plot is divided into eight bins corresponding to equal intervals of width $\pi/4$ in the strong-phase difference $\Delta\delta_D$ predicted by an amplitude model~\cite{CLEO:KShhStrongPhase}. The bin integrals and local strong-phase difference are defined by
\begin{equation}
\begin{aligned}
K_i&=\int_i \left|A(\Phi)\right|^2\mathrm{d}\Phi,\\
K_{-i}&=\int_i \left|\bar{A}(\Phi)\right|^2\mathrm{d}\Phi,\\
\Delta\delta_D(\Phi)&=\arg\!\left[A(\Phi)\bar{A}^{*}(\Phi)\right],\\
c_i+\mathrm{i}s_i&=
\frac{\int_i A(\Phi)\bar{A}^{*}(\Phi)\mathrm{d}\Phi}
{\sqrt{K_iK_{-i}}}.
\end{aligned}
\label{eq:kspipi-strong-phase}
\end{equation}
The parameters $c_i$ and $s_i$ are the amplitude-weighted averages of $\cos\Delta\delta_D$ and $\sin\Delta\delta_D$ over each dalitz bin.
In the $D\to K_S^0\pi^+\pi^-$ tag mode, the bin parameters are related to the quantities entering Eq.~\eqref{eq:time-integrated-DT-rates} by
\begin{equation}
\begin{aligned}
P_g^i&=K_i,\qquad \bar P_g^i=K_{-i},\\
Z_g^i&=\sqrt{K_iK_{-i}}\left(c_i+\mathrm{i}s_i\right).
\end{aligned}
\label{eq:kspipi-PZ}
\end{equation}
Positive bins lie in the region $m_+^2>m_-^2$, where $m_\pm^2=m^2(K_S^0\pi^\pm)$. Bin $-i$ is obtained from bin $i$ by interchanging $m_+^2$ and $m_-^2$. The hadronic inputs for this mode are taken from the BESIII measurement and fixed to their central values in the fit~\cite{BESIII:KSpipiStrongPhase}.

The expressions for the time-integrated DT decay rates show that the sensitivity to the hadronic and mixing parameters depends on both the tag category and the $C$ eigenvalue. The flavour, $C\!P$, and self-conjugate tags probe different combinations of the hadronic and mixing terms, while the $C$-odd and $C$-even configurations differ in the signs of the interference terms and the coefficients of the mixing contributions. Combining the corresponding DT decay rates therefore allows the hadronic parameters to be determined simultaneously with $x$, $y$, $|q/p|$, and $\phi$.

\section{Monte Carlo simulation and event selection}
\label{sec:simulation-selection}

\subsection{Monte Carlo simulation}

The STCF detector is a general-purpose detector designed for $e^+e^-$ collisions. It consists of a tracking system with inner and outer trackers, a particle-identification (PID) system, an electromagnetic calorimeter that provides good position resolution for photons and electrons, and a muon detector that provides good separation between muons and pions. A detailed description of each detector system is given in Ref.~\cite{STCF:CDR}. 
The detector response is modeled using the OSCAR full-simulation framework~\cite{STCF:OSCAR}. The simulation includes the detailed detector geometry and materials, as well as particle interactions with the detector components. The simulated detector signals are digitized and processed with the standard OSCAR reconstruction algorithms. The reconstructed charged tracks, photon candidates, and PID information are then used in the event selection described below.

Monte Carlo (MC) samples are generated at $\sqrt{s}=3.773$ and $4.030$~GeV. Initial-state radiation and the beam-energy spread are modeled with \textsc{kkmc}~\cite{Jadach:KKMC,Jadach:CEEX}, while the subsequent decays are simulated with \textsc{evtgen}~\cite{Lange:EvtGen,Ping:EvtGenBESIII} using branching fractions from the Particle Data Group~\cite{PDG:2026}. Decay modes without measured branching fractions are described by the \textsc{lundcharm} model~\cite{Chen:LundCharm}, and final-state radiation is simulated with \textsc{photos}~\cite{Barberio:PHOTOS}. At $\sqrt{s}=3.773$~GeV, the simulated sample consists of $\epem\to\psi(3770)\to D^0\bar{D}^0$ events. The sample at $\sqrt{s}=4.030$~GeV includes the processes $\epem\to D^{(*)0}\bar{D}^{(*)0}$, followed by $D^{*0}\to\gamma D^0$ or $D^{*0}\to\pi^0D^0$. Quantum correlations of the $D^0\bar{D}^0$ pairs are taken into account in the MC simulation at both energies. Large signal MC samples are used to determine the efficiency for each signal--tag combination. Together with the time-integrated rates in Eq.~\eqref{eq:time-integrated-DT-rates}, these efficiencies are used to generate ensembles of pseudodata corresponding to an integrated luminosity of $1~\mathrm{ab}^{-1}$ at each energy. 

\subsection{Event selection}

The basic event-selection requirements follow those used in the corresponding BESIII quantum-correlated $D\bar{D}$ analyses~\cite{BESIII:K3piKpipi0:2021,8fb-1}. Charged tracks are required to satisfy $|\cos\theta|<0.93$, where $\theta$ is measured with respect to the $z$ axis along the beam direction. Except for tracks used to reconstruct $K_S^0$ mesons, the distance of closest approach to the interaction point (IP) is required to satisfy $|V_z|<10$~cm along the $z$ axis and $|V_{xy}|<1$~cm in the transverse plane.
PID for charged tracks combines measurements of the specific ionization ($dE/dx$) and the flight time provided by the time-of-flight (TOF) system to construct likelihoods $\mathcal{L}(h)$ for the hypotheses $h=K,\pi$. A track is identified as a kaon if $\mathcal{L}(K)>\mathcal{L}(\pi)$ and as a pion if $\mathcal{L}(\pi)>\mathcal{L}(K)$.
Photon candidates are selected from isolated showers in the electromagnetic calorimeter (EMC). The deposited energy is required to exceed $25$~MeV in the barrel region, $|\cos\theta|<0.80$, or $50$~MeV in the end-cap region, $0.86<|\cos\theta|<0.92$. To suppress showers originating from charged tracks, an EMC shower must be separated by more than $10^{\circ}$ from the nearest charged track extrapolated to the EMC.

The $K_S^0$ candidates are formed from pairs of oppositely charged tracks, with no particle-identification requirement imposed. The two tracks are fitted to a common vertex, and their invariant mass is required to satisfy $0.487<M_{\pi^+\pi^-}<0.511$~GeV/$c^2$. The decay length of the $K_S^0$ candidate is required to exceed twice the vertex resolution away from the IP. The $\pi^0$ and $\eta$ candidates are formed from photon pairs with invariant masses in the ranges $0.115<M_{\gamma\gamma}<0.150$~GeV/$c^2$ and $0.480<M_{\gamma\gamma}<0.580$~GeV/$c^2$, respectively. A mass-constrained fit is then applied to each $\pi^0$ or $\eta$ candidate. The $\omega$ and $\eta'$ candidates are reconstructed in the $\pi^+\pi^-\pi^0$ and $\gamma\pi^+\pi^-$ final states, respectively. The $\omega$ candidates must satisfy $0.750<M_{\pi^+\pi^-\pi^0}<0.820$~GeV/$c^2$, while $0.940<M_{\gamma\pi^+\pi^-}<0.970$~GeV/$c^2$ is imposed for the $\eta'$ candidates.

\begin{figure}[tb]
\centering
\includegraphics[width=0.8\linewidth]{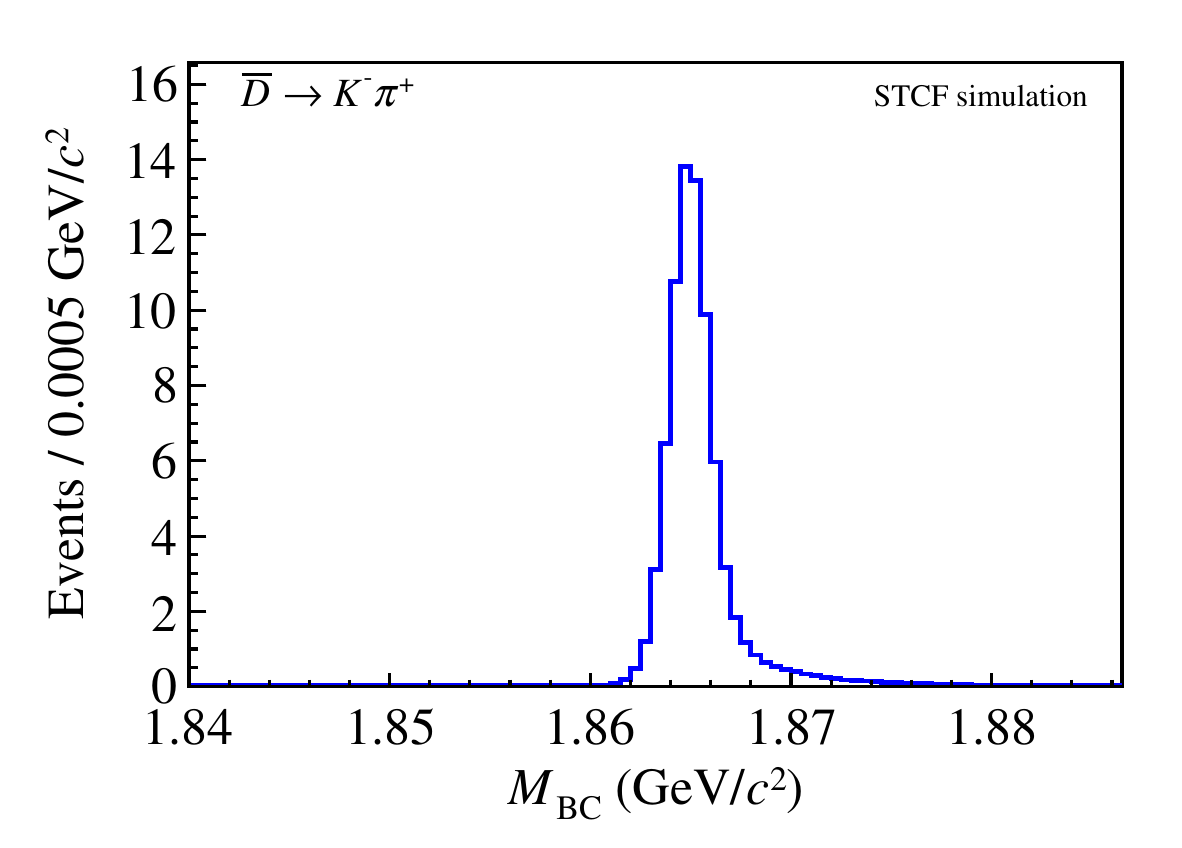}
\caption{The $M_{\rm BC}$ distribution for the $D\to K^-\pi^+$ and $\bar{D}\to K^-\pi^+$ LS DT combination in the signal MC sample.}
\label{fig:kpi-ls-mbc}
\end{figure}

Neutral-$D$ candidates are reconstructed in the signal modes $D\to K^-\pi^+\pi^+\pi^-$, $K^-\pi^+\pi^0$, and $K^-\pi^+$, as well as in the tag modes listed in Table~\ref{tab:tag-modes}. 
Backgrounds involving $D\to K_S^0\to\pip\pim$ decays in
$D\to\km\pip\pim\pip$ and $\pip\pim\piz$ are suppressed by requiring
$M_{\pip\pim}$ to lie outside $(0.48,0.51)\,\mathrm{GeV}/c^2$.
The selection of signal candidates at $\sqrt{s}=3.773$~GeV is based on the energy difference $\Delta E=E_D-E_{\rm beam}$ and the beam-constrained mass $M_{\rm BC}=\sqrt{E_{\rm beam}^{2}-|\boldsymbol{p}_D|^{2}}$, where $E_D$ and $\boldsymbol{p}_D$ are the reconstructed energy and momentum of the $D$ candidate in the center-of-mass frame. The $\Delta E$ selection interval for each tag mode is chosen to be the same as that used in the corresponding BESIII analysis~\cite{8fb-1}.
The $M_{\rm BC}$ distributions of correctly reconstructed candidates peak near the nominal $D^0$ mass. As a representative example, the distribution for $D \to K^-\pi^+$ is shown in Fig.~\ref{fig:kpi-ls-mbc}.
In the corresponding BESIII analysis with similar selection criteria, the $M_{\rm BC}$ distributions of the flavour-, $C\!P$-, and $K_S^0\pi^+\pi^-$-tagged samples contain combinatorial backgrounds from $D\bar{D}$ events and continuum $e^+e^-\to q\bar{q}$ production. In the LS flavour-tagged samples, OS events can form peaking backgrounds due to particle misidentification. Peaking backgrounds from other $D$ decays are also present in the $C\!P$-tagged samples, with generally small contributions~\cite{8fb-1}.

The reconstruction strategy at $\sqrt{s}=4.030$~GeV differs from that at $\sqrt{s}=3.773$~GeV due to the presence of transition photons or $\pi^0$ mesons from $D^{*0}$ decays. The two $D$ candidates are first subjected to kinematic fits in which the masses of the intermediate states and the $D^0$ mesons are constrained to their known values. If more than one $D^0\bar{D}^0$ combination is retained, the candidate with the smallest sum of the fit $\chi^2$ values is selected. Three complementary partial-reconstruction methods are then used, referred to as the $D\bar{D}$, $\gamma D\bar{D}$, and $\pi^0D\bar{D}$ methods. Only the two $D$ mesons are reconstructed in the $D\bar{D}$ method. The $\gamma D\bar{D}$ and $\pi^0D\bar{D}$ methods additionally include a reconstructed transition photon and $\pi^0$, respectively.

\begin{figure}[tb]
\centering
\includegraphics[width=0.8\linewidth]{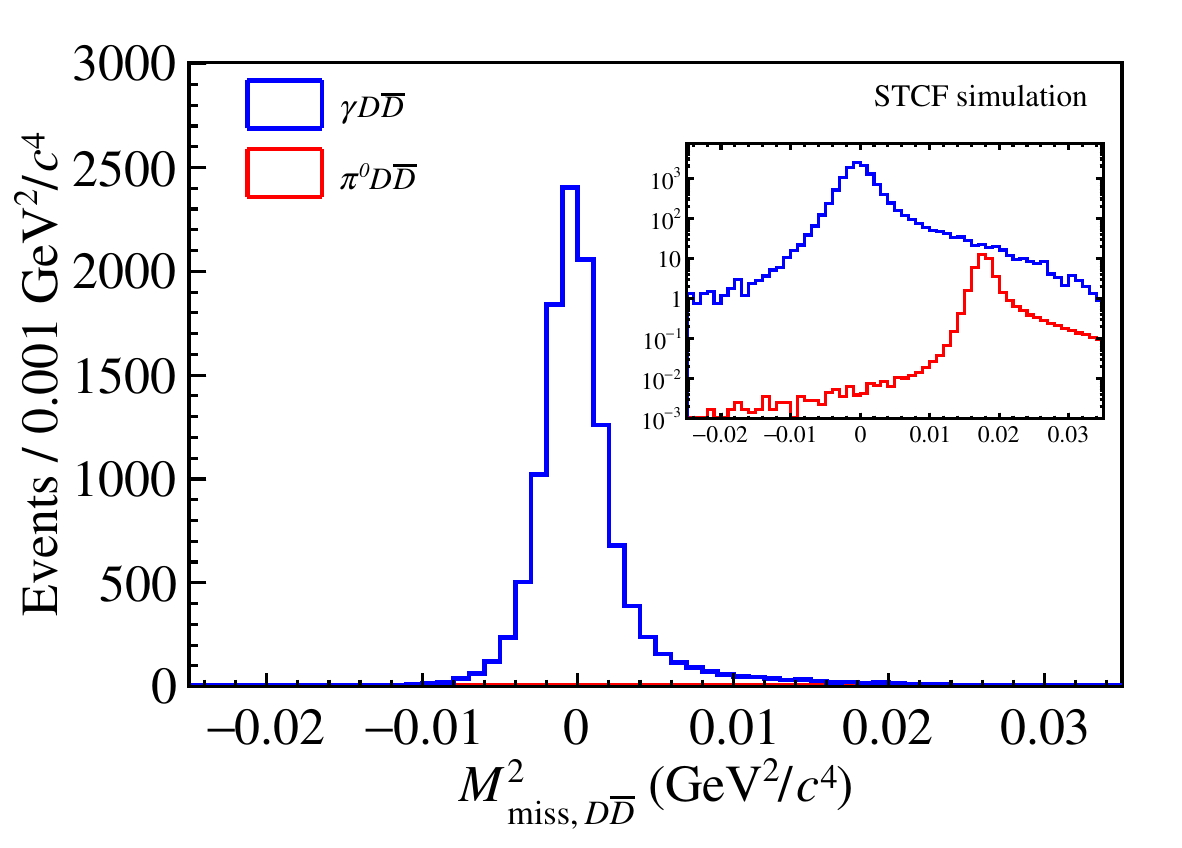}
\caption{The $M_{{\rm miss},D\bar{D}}^2$ distribution for the $D\to K^-\pi^+$ and $\bar{D}\to K^-\pi^+$ LS DT combination in the signal MC sample. The blue and red histograms correspond to $D^{*0}\to\gamma D^0$ and $D^{*0}\to\pi^0D^0$, respectively. The inset shows the same distributions on a logarithmic vertical scale to display the suppressed $\pi^0D\bar{D}$ contribution.}
\label{fig:kpi-ls-mmiss2}
\end{figure}

The $D\bar{D}$ method is used for the $\epem\to D^{*0}\bar{D}^0$ process, and the missing energy and momentum are calculated from the reconstructed $D\bar{D}$ pair as
\begin{equation}
\begin{aligned}
E_{{\rm miss},D\bar{D}} &= E_{\rm cm}-E_f-E_g,\\
\boldsymbol{p}_{{\rm miss},D\bar{D}} &= \boldsymbol{p}_{\rm cm}-\boldsymbol{p}_f-\boldsymbol{p}_g.
\end{aligned}
\end{equation}
Here, $(E_{\rm cm},\boldsymbol{p}_{\rm cm})$ is the four-momentum of the initial $\epem$ system, with $\boldsymbol{p}_{\rm cm}=0$ in the center-of-mass frame.
The corresponding missing mass is defined as
\begin{equation}
M_{{\rm miss},D\bar{D}}=\sqrt{E_{{\rm miss},D\bar{D}}^2-\left|\boldsymbol{p}_{{\rm miss},D\bar{D}}\right|^2}.
\end{equation}
As an example, Fig.~\ref{fig:kpi-ls-mmiss2} shows the $M_{{\rm miss},D\bar{D}}^2$ distribution for the LS DT combination $(K^-\pi^+,K^-\pi^+)$, with charge-conjugate modes implied. The structures near zero and $m_{\pi^0}^2$ correspond to the $D^{*0}\bar{D}^0\to\gamma D^0\bar{D}^0$ and $D^{*0}\bar{D}^0\to\pi^0D^0\bar{D}^0$ processes, which produce $C$-even and $C$-odd $D^0\bar{D}^0$ pairs, respectively, and allow the two contributions to be separated.
The strong suppression of the $\pi^0D^0\bar{D}^0$ contribution in this LS sample arises from destructive quantum interference in the $C$-odd state. For the identical two-body final states $K^-\pi^+$, the coherence factor is unity, and the two amplitudes involving one Cabibbo-favored and one doubly Cabibbo-suppressed decay cancel exactly in the absence of charm mixing. The remaining $C$-odd DT rate therefore starts at order $x^2+y^2$, as follows from Eq.~\eqref{eq:time-integrated-DT-rates}. For the $C$-even state, these amplitudes add constructively, resulting in a much larger $\gamma D^0\bar{D}^0$ contribution.

Events from $\epem\to D^{*0}\bar{D}^{*0}$ production are reconstructed with the $\gamma D\bar{D}$ and $\pi^0D\bar{D}$ methods. Photon and $\pi^0$ candidates are formed from showers not assigned to either $D$ decay. Each candidate is then combined with the two $D$ candidates in turn to form a $\gamma D$ or $\pi^0D$ candidate. For each combination, a kinematic fit constrains the $\gamma D$ or $\pi^0D$ system and its recoil system to the nominal $D^{*0}$ mass. The combination with the smaller $\chi^2$ is retained, subject to $\chi^2<80$. The missing-mass-squared variables $M_{{\rm miss},\gamma D\bar{D}}^2$ and $M_{{\rm miss},\pi^0D\bar{D}}^2$ are defined analogously to $M_{{\rm miss},D\bar{D}}^2$ in the $D\bar{D}$ method, with the four-momentum of the reconstructed transition photon or $\pi^0$ additionally subtracted. Their distributions identify the unreconstructed transition particle and distinguish among the $\gamma\pi^0D^0\bar{D}^0$, $\gamma\gamma D^0\bar{D}^0$, and $\pi^0\pi^0D^0\bar{D}^0$ final states, thereby providing samples of $C$-even and $C$-odd $D^0\bar{D}^0$ pairs.

Related BESIII studies use a similar partial-reconstruction method~\cite{dkgg-9twj,q6qr-8csr}. The backgrounds include continuum $e^+e^-\to q\bar{q}$ production, incorrectly reconstructed $D$ decays, and processes involving charged $D^{*}$ mesons. Initial-state radiation also causes cross feed among the production mechanisms.

\begin{table*}[!t]
    \caption{Expected statistical precisions of the $D$-decay hadronic parameters at STCF and the current measurements.}
    \label{tab:hadronic-parameter-precision}
    \centering
    \renewcommand{\arraystretch}{1.15}
    \setlength{\tabcolsep}{8pt}
    \begin{tabular}{lcccc}
        \hline\hline
        \multirow{2}{*}{Parameter} & \multicolumn{2}{c}{STCF expected precision} & \multirow{2}{*}{Current measurement} & \multirow{2}{*}{Reference} \\
        \cline{2-3}
        & $3.773$~GeV only & $3.773$ and $4.030$~GeV & & \\
        \hline
        $r_{K3\pi}^{1}$ ($\%$)          & 0.021 & 0.015                   & $5.48\pm0.05$             & \multirow{12}{*}{Ref.~\cite{8fb-1}}       \\
        $r_{K3\pi}^{2}$ ($\%$)          & 0.023 & 0.016                   & $5.96\pm0.05$             &                                           \\
        $r_{K3\pi}^{3}$ ($\%$)          & 0.022 & 0.016                   & $5.86\pm0.05$             &                                           \\
        $r_{K3\pi}^{4}$ ($\%$)          & 0.018 & 0.013                   & $4.83\pm0.04$             &                                           \\
        $\delta_{K3\pi}^{1}$ ($^\circ$) & 0.76 & 0.60                    & $124^{+10}_{-8}$          &                                           \\
        $\delta_{K3\pi}^{2}$ ($^\circ$) & 1.5 & 1.2                     & $170^{+14}_{-12}$         &                                           \\
        $\delta_{K3\pi}^{3}$ ($^\circ$) & 1.2 & 0.94                     & $179\pm11$                &                                           \\
        $\delta_{K3\pi}^{4}$ ($^\circ$) & 0.76 & 0.55                    & $260^{+10}_{-13}$         &                                           \\
        $R_{K3\pi}^{1}$                 & 0.013 & 0.010                   & $0.72\pm0.14$             &                                           \\
        $R_{K3\pi}^{2}$                 & 0.0068 & 0.0050                  & $0.65^{+0.08}_{-0.07}$    &                                           \\
        $R_{K3\pi}^{3}$                 & 0.0049 & 0.0039                  & $0.80^{+0.07}_{-0.06}$    &                                           \\
        $R_{K3\pi}^{4}$                 & 0.016 & 0.012                   & $0.54^{+0.18}_{-0.19}$    &                                           \\
        \hline
        $r_{K\pi\pi^0}$ ($\%$)            & 0.022 & 0.016                   & $4.41\pm0.08$             & \multirow{3}{*}{Ref.~\cite{8fb-1}}        \\
        $\delta_{K\pi\pi^0}$ ($^\circ$)   & 0.65 & 0.50                    & $209^{+7}_{-8}$           &                                           \\
        $R_{K\pi\pi^0}$                   & 0.0030 & 0.0026                  & $0.75\pm0.03$             &                                           \\
        \hline
        $r_{K\pi}$ ($\%$)                 & 0.023 & 0.018                   & $5.855^{+0.009}_{-0.010}$ & \multirow{2}{*}{Ref.~\cite{LHCb:2024yxi}} \\
        $\delta_{K\pi}$ ($^\circ$)        & 0.58 & 0.44                    & $191.4\pm2.4$             &                                           \\
        \hline\hline
    \end{tabular}
\end{table*}

\section{Fit procedure}
\label{sec:fit-procedure}

A simultaneous $\chi^2$ fit is performed using the flavour-tagged, $C\!P$-tagged, and self-conjugate-tagged observables. 
For each combination $(f_i,g_j)$, the expected signal yield is
\begin{equation}
N_{ij}^{C}(\boldsymbol{\theta})
=N_{D\bar D}^{C}\,\epsilon_{C,ij}\,
\mathcal B_C(f_i,g_j;\boldsymbol{\theta}),
\label{eq:expected-DT-yields}
\end{equation}
where $N_{D\bar D}^{C}=\mathcal L\,\sigma_{D\bar D}^{C}$ is the number of produced $D^0\bar D^0$ pairs with charge-conjugation eigenvalue $C$.
Here, $\mathcal L$ is the integrated luminosity and $\sigma_{D\bar D}^{C}$ is the effective production cross section, including the relevant $D^{*0}$ decay branching fractions for channels involving $D^{*0}$ production.
The quantity $\epsilon_{C,ij}$ denotes the DT efficiency determined from simulation.
The joint decay probability $\mathcal B_C(f_i,g_j;\boldsymbol{\theta})$ is calculated from the time-integrated rates in Eq.~\eqref{eq:time-integrated-DT-rates}.

The parameters allowed to vary in the fit are
\begin{equation}
\boldsymbol{\theta}=
\left(
x, y, \left|q/p\right|, \phi, r^i_f,\delta^i_f, R^i_{f}
\right),
\label{eq:fit-parameters}
\end{equation}
where $f=K^-\pi^+\pi^+\pi^-$, $K^-\pi^+\pi^0$, and $K^-\pi^+$. These parameters include the signal-side hadronic parameters, the mixing parameters $x$ and $y$, and the indirect-$C\!P$-violation parameters $|q/p|$ and $\phi$. For the two-body $K^-\pi^+$ mode, the coherence factor is fixed to one by definition.

The function minimized in the fit is written as
\begin{equation}
\chi^2(\boldsymbol{\theta})=
\chi^2_{\rm LS}+\chi^2_{\rm OS}+\chi^2_{C\!P}
+\chi^2_{K_S^0\pi^+\pi^-},
\label{eq:total-chi2}
\end{equation}
where the four terms are from the LS, OS, $C\!P$, and $K_S^0\pi^+\pi^-$ tags. For each tag group $k$, the $\chi^2$ term is
\begin{equation}
\chi^2_k=
\left[\boldsymbol{O}_k-\boldsymbol{P}_k(\boldsymbol{\theta})\right]^{\mathrm T}
V_k^{-1}
\left[\boldsymbol{O}_k-\boldsymbol{P}_k(\boldsymbol{\theta})\right],
\label{eq:category-chi2}
\end{equation}
where $\boldsymbol{O}_k$ is the vector of pseudodata observables, $\boldsymbol{P}_k$ is the vector of predicted values, and $V_k$ is the statistical covariance matrix.

The loss of statistical precision due to background contamination and signal extraction is estimated using the signal yields and their statistical uncertainties reported by BESIII~\cite{8fb-1,dkgg-9twj,q6qr-8csr}.
For each measured channel, we define an uncertainty scaling factor $\kappa=\sigma_{\mathrm{BESIII}}/\sqrt{N_{\mathrm{sig}}}$, where $N_{\mathrm{sig}}$ is the signal yield before efficiency corrections and $\sigma_{\mathrm{BESIII}}$ is its statistical uncertainty.
For channels without direct measurements, $\kappa$ is estimated from related decay modes in the same production and $C$-parity category.
The statistical uncertainties entering the sensitivity fit are multiplied by these factors, while the expected signal yields remain unchanged.

For each pseudoexperiment, all free parameters are determined in one fit. The uncertainty of one parameter is defined by $\Delta\chi^2=1$ after minimizing over the other free parameters.

\section{Expected sensitivities}
\label{sec:expected-sensitivities}

\subsection{Hadronic parameters}
The expected statistical uncertainties on the hadronic parameters at STCF are summarized in Table~\ref{tab:hadronic-parameter-precision}, together with the current measurements. 
Based on the STCF simulation, these uncertainties are evaluated using a fit at $\sqrt{s}=3.773$ GeV alone and a simultaneous fit at $\sqrt{s}=3.773$ and $4.030$ GeV.
The largest improvement in precision is seen for the strong-phase differences and coherence factors of the multibody modes. More precise knowledge of these quantities improves the description of the interference between the Cabibbo-favored and doubly Cabibbo-suppressed amplitudes. It therefore reduces the uncertainties associated with external hadronic inputs in measurements of the global charm mixing and in determinations of the CKM angle $\gamma$ from $B\to DK$ decays.

In the published LHCb measurement of $\gamma$ using $B^{\pm}\to D[K^{\mp}\pi^{\pm}\pi^{\pm}\pi^{\mp}]h^{\pm}$ ($h=K,\pi$) decays~\cite{LHCb:GammaK3pi:2023}, the uncertainty from the external strong-phase differences and coherence factors is already comparable to the statistical uncertainty.
The effect of the projected STCF hadronic-parameter inputs on a future determination of $\gamma$ is evaluated using the binned phase-space method adopted in the LHCb analysis. The decay-rate observables and their uncertainties measured with this $9~\mathrm{fb}^{-1}$ Run~1 and Run~2 data sample are first included in a binned $\chi^2$ fit. The fitted values of $\gamma$ and the $B$-decay parameters are then used to construct a pseudodata sample corresponding to $300~\mathrm{fb}^{-1}$, taken as a benchmark for the LHCb Upgrade II data sample. For $B^{\pm}\to D[K^{\mp}\pi^{\pm}\pi^{\pm}\pi^{\mp}]K^{\pm}$ decays, the projected statistical uncertainty on $\gamma$ is approximately $1^\circ$ when the $D$-decay parameters are assumed to be precisely known~\cite{LHCb:UpgradeII}. The statistical uncertainties on the projected STCF hadronic parameters contribute $0.52^\circ$ to the uncertainty on $\gamma$ in this channel. This contribution is smaller than the projected LHCb statistical uncertainty, suggesting that STCF could help reduce the uncertainty associated with the hadronic inputs in this measurement.

\subsection{Charm mixing and indirect $C\!P$ violation}
\label{subsec:mixing-cpv}
In addition to the hadronic parameters discussed above, the simultaneous fit provides sensitivity to the charm-mixing parameters $x$ and $y$, and to the indirect-$C\!P$-violation parameters $|q/p|$ and $\phi$. 
The sensitivity to $x$ and $y$ arises mainly from terms that are linear in these parameters in the time-integrated rates of the $C$-even $\DDbar$ sample.
Since both $C$-even and $C$-odd samples are available at $\sqrt{s}=4.030$~GeV, this data sample plays an important role in the simultaneous fit.
Combining the two samples allows charm mixing to be studied without using decay-time information and enables the determination of the hadronic parameters.

\begin{figure}[tb]
\centering
\includegraphics[width=0.8\linewidth]{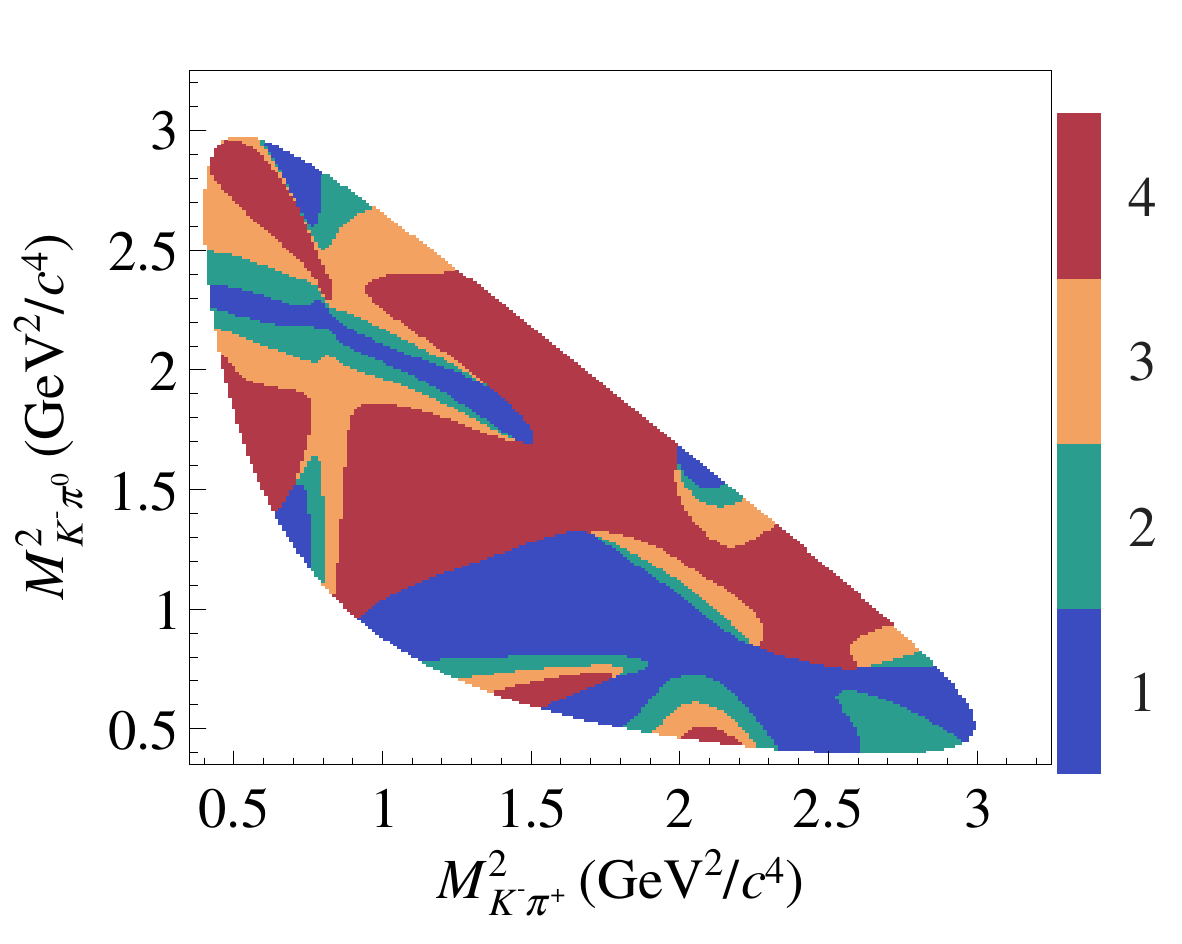}
\caption{Four-bin Dalitz-plot scheme for $D\to K^-\pi^+\pi^0$. The colours show the bin number assigned according to the local strong-phase difference between the Cabibbo-favored and doubly Cabibbo-suppressed amplitudes.}
\label{fig:kpipi0-dalitz-bins}
\end{figure}

For the decay $D\to K^-\pi^+\pi^0$, both the integrated approach and a four-bin approach are studied.
The unbinned approach uses observables integrated over the full phase space, with a single average strong-phase difference and a single coherence factor.
The four-bin approach uses a binning scheme defined by the local strong-phase difference $\Delta\delta_D(\Phi)$.
For the bin assignment, $\Delta\delta_D(\Phi)$ is mapped to the range $[-180^\circ,180^\circ]$ and divided into four intervals: $[-180^\circ,-21^\circ]$, $[-21^\circ,-1^\circ]$, $[-1^\circ,22^\circ]$, and $[22^\circ,180^\circ]$. The bin boundaries are chosen to group regions with similar strong-phase differences while keeping the event yields in the four bins approximately balanced. This choice retains sufficient statistics in each bin and reduces cancellations of the interference terms over phase space.
The resulting four-bin partition of the $D\to K^-\pi^+\pi^0$ Dalitz plot is shown in Fig.~\ref{fig:kpipi0-dalitz-bins}, using the squared invariant masses $M_{K^-\pi^+}^2$ and $M_{K^-\pi^0}^2$ as the kinematic variables. 
A single bin can contain several separate regions because the assignment is based on the local strong-phase difference rather than on rectangular boundaries in the Dalitz plot.

The expected statistical uncertainties for the integrated and four-bin approaches are compared in Table~\ref{tab:mixing-cpv-precision}. The four-bin approach provides better sensitivity to both the charm-mixing and indirect-$C\!P$-violation parameters. This improvement arises from retaining the variation of the strong-phase difference across the $D\to K^-\pi^+\pi^0$ Dalitz plot rather than from a larger sample size, as the same total signal yield is used in both approaches. These results demonstrate that time-integrated threshold measurements at STCF can provide information complementary to that obtained from decay-time-dependent studies of charm mixing and indirect $C\!P$ violation.

\begin{table}[tb]
\caption{Expected statistical precision for the charm-mixing and indirect-$C\!P$-violation parameters at STCF and the current measurements.}
\label{tab:mixing-cpv-precision}
\centering
\small
\renewcommand{\arraystretch}{1.25}
\setlength{\tabcolsep}{4pt}
\begin{tabular}{@{}lccc@{}}
\hline\hline
\multirow{2}{*}{Parameter} & \multicolumn{2}{c}{STCF expected uncertainty} & HFLAV 2025 \\
\cline{2-3}
 & Integrated & Four-bin & Result \\
\hline
$x$ ($\times10^{-3}$) & 0.53 & 0.43 & $4.05\pm0.43$ \\
$y$ ($\times10^{-3}$) & 0.26 & 0.23 & $6.36\pm0.24$ \\
$|q/p|$          & 0.029 & 0.027 & $0.983^{+0.015}_{-0.014}$ \\
$\phi$ ($^\circ$) & 2.0 & 1.9 & $-1.51^{+1.03}_{-1.06}$ \\
\hline\hline
\end{tabular}
\end{table}

\section{Summary}
\label{sec:summary}

In summary, we evaluate the expected sensitivities to charm mixing, indirect $C\!P$ violation, and hadronic parameters using the quantum-correlated $D^0\bar{D}^0$ pairs expected at STCF at $\sqrt{s}=3.773$ and $4.030$~GeV. The expected signal yields for the $C$-even and $C$-odd states are combined in a simultaneous fit to determine the charm mixing parameters $x$ and $y$, the indirect $C\!P$ violation parameters $|q/p|$ and $\phi$, as well as the hadronic parameters of the $D\to K^-\pi^+$, $D\to K^-\pi^+\pi^0$, and $D\to K^-\pi^+\pi^+\pi^-$ decays. The different interference patterns in the two quantum-correlation cases allow these paramters to be extracted without using decay-time information.
The studies based on the STCF simulation show improved precision on the strong-phase differences and coherence factors of the $D$ multibody decays. A four-bin Dalitz-plot approach based on the local strong-phase difference is also studied for the $D\to K^-\pi^+\pi^0$ decay. By accounting for the phase variation across the Dalitz plot, the four-bin approach reduces cancellations of the interference terms and provides better sensitivity to charm mixing and indirect $C\!P$ violation than the phase space integrated approach under the same statistics.
The STCF determinations of charm mixing and indirect $C\!P$ violation would complement time-dependent measurements at Belle II and LHCb. In addition, the hadronic-parameter measurements would provide external inputs to the global charm-mixing analyses and to the CKM angle $\gamma$ measurement from $B\to DK$ decays. 
The expected statistical uncertainties at STCF on the strong-phase differences and coherence factors for $D\to K^-\pi^+\pi^+\pi^-$ decay are $0.56^\circ$–$1.3^\circ$ and $0.0039$–$0.013$, respectively, across the four phase-space bins.
A projected binned analysis of $B^{\pm}\to D[K^{\mp}\pi^{\pm}\pi^{\pm}\pi^{\mp}]K^{\pm}$ decays indicates that the statistical uncertainties on the STCF hadronic inputs for this $D$ decay contribute $0.52^\circ$ to the uncertainty on $\gamma$, demonstrating their value for future precision measurements.

\begin{acknowledgments}
We thank the supercomputing center of Lanzhou university for their strong support.
This work is supported by the Fundamental Research Funds for the Central Universities, Lanzhou University under Contracts No.~lzujbky-2025-ytB01 and No.~lzujbky-2023-stlt01; the National Key R\&D Program of China under Contracts No. 2022YFA1602200 and No. 2023YFA1607200; the National
Natural Science Foundation of China (NSFC) under Contracts No. 12405112, No. 12422504, No. 12341501, No. 12341503, No. 12341504 and No. 12221005; the international partnership program of the Chinese Academy of Sciences Grant No. 211134KYSB20200057; the Excellent Youth Foundation of Hunan Provincial Department of Education under Contracts No. 23B0429. We thank the Hefei Comprehensive National Science Center for their strong support on the STCF key technology research project.
\end{acknowledgments}

%
%


\bibliographystyle{apsrev4-2}
\bibliography{apssamp}

\end{document}